\documentclass[pdflatex,sn-mathphys-num]{sn-jnl}
\usepackage{amsmath,amssymb,amsfonts}
\usepackage{bm}
\usepackage{graphicx}
\usepackage{enumitem}
\usepackage{booktabs}
\usepackage{array}
\usepackage{geometry}
\usepackage{hyperref}

\hypersetup{colorlinks=true,linkcolor=blue,citecolor=blue,urlcolor=blue}

\newcommand{\Lcal}{\mathcal{L}}
\newcommand{\Hcal}{\mathcal{H}}
\newcommand{\Vhat}{\hat V}
\newcommand{\Heff}{\Hcal_{\rm eff}}
\newcommand{\Dvec}{\mathbf{D}}
\newcommand{\Hvec}{\mathbf{H}}
\newcommand{\Evec}{\mathbf{E}}
\newcommand{\Bvec}{\mathbf{B}}
\newcommand{\nvec}{\mathbf{n}}
\newcommand{\dd}{\mathrm{d}}

\newcommand{\kerop}{\mathrm{Ker}}

\title[Constitutive Origin of Hamiltonian Degeneracy]
{Constitutive Origin of Hamiltonian Degeneracy in Nonlinear Electrodynamics with Spontaneous Lorentz Symmetry Breaking}

\author*[1]{\fnm{C. A.} \sur{Escobar}}\email{carlos.escobar@xanum.uam.mx}

\author*[1]{\fnm{Rom\'an} \sur{Linares}}\email{lirr@xanum.uam.mx}

\affil*[1]{
\orgdiv{Departamento de F\'isica},
\orgname{Universidad Aut\'onoma Metropolitana-Iztapalapa},
\orgaddress{
\street{San Rafael Atlixco 186},
\city{Ciudad de M\'exico},
\postcode{09340},
\country{M\'exico}
}
}

\abstract{
In Pleba\'nski nonlinear electrodynamics with spontaneous Lorentz symmetry breaking,
nontrivial magnetic backgrounds are selected by stationary points of an effective Hamiltonian.
Previous branchwise Hamiltonian analyses showed that this same stationarity requirement
coincides with the vanishing of the determinant of the Poisson-bracket matrix among the
second-class constraints, but the structural origin of this coincidence was not manifest.
We show that it follows from the constitutive origin of the theory. The structural potential
\(V(P,Q)\) generates the electromagnetic constitutive relations, while the effective
Hamiltonian for magnetic vacua is the complementary energy associated with the magnetic
response at fixed \(\Dvec\). Moreover, because the first-order constitutive relation enters
the Dirac constraint structure, the magnetic constitutive Jacobian appears as a local block
of the Poisson-bracket matrix among the second-class constraints. This complementary-energy
structure implies that every nontrivial magnetic stationary point lies on a surface where
the linearized map \(\delta\Hvec\mapsto\delta\Bvec\), at fixed \(\Dvec\), loses rank.
We use this interpretation to formulate the reduced linearized theory at the vacuum,
discuss the removal of the radial mode in the vacuum-restricted theory, and clarify why
electric and mixed stationary branches are obstructed in single-invariant models.
}


\begin{document}

\maketitle

\section{Introduction}

Lorentz invariance is a central structural principle of both the Standard Model and
general relativity. Nevertheless, several approaches to physics beyond the Standard Model
allow for the possibility that Lorentz symmetry may be broken, either explicitly or
spontaneously~\cite{KosteleckySamuel1989,ColladayKostelecky1997,ColladayKostelecky1998}. In
spontaneous Lorentz symmetry breaking, tensorial fields acquire nonzero vacuum
expectation values, so that the vacuum preserves only the subgroup that leaves the
background invariant. Familiar examples include vector models with nonzero vacuum
expectation values and scalar derivative backgrounds, such as bumblebee and
ghost-condensate models~\cite{KosteleckySamuel1989,BluhmKostelecky2005,BluhmGagnePottingVrublevskis2008,
ArkaniHamedGhostCondensate}.
Hamiltonian subtleties in Lorentz-breaking vector models are well known. In
bumblebee-type theories, the existence of a nonzero minimum of a Lorentz-breaking
potential is not, by itself, sufficient to characterize a physically admissible vacuum.
The Hamiltonian structure, constraints, stability properties, and effective number of
degrees of freedom must also be analyzed
~\cite{BluhmGagnePottingVrublevskis2008,EscobarLinares2022,ZhuLiXiao2026}.
In this sense, bumblebee models provide an important precedent: Lorentz-breaking
vacua in constrained nonlinear field theories should be characterized at the Hamiltonian
level, not merely as minima of a Lagrangian potential.

The Pleba\'nski nonlinear electrodynamics (NLED) models considered below are not bumblebee theories: their
Lorentz-breaking order parameter belongs to the field-strength/constitutive sector, rather
than to a fundamental vector field governed by a nonderivative Lorentz-breaking potential.
They nevertheless realize the same Hamiltonian principle in a gauge-invariant setting.
In the Pleba\'nski NLED framework, nontrivial electromagnetic vacua, Hamiltonian
consistency, and constitutive response are naturally intertwined. In the first-order
formulation introduced by Pleba\'nski, the antisymmetric tensor \(P^{\mu\nu}\) and the
gauge potential \(A_\mu\) are treated as independent variables, and the nonlinear
electromagnetic response is encoded in a structural potential \(V(P,Q)\). This formulation
is particularly useful for constrained Hamiltonian analysis, because the natural variables
are the analogues of electric displacement and magnetic intensity, \(\Dvec\) and \(\Hvec\),
rather than \(\Evec\) and \(\Bvec\) themselves
~\cite{Plebanski1970,Dirac1964,HenneauxTeitelboim1992}. More broadly, nonlinear electrodynamics has also been studied in connection with
duality, characteristic propagation, optical metrics, and causality or energy conditions
~\cite{Boillat1970,GibbonsRasheed1995,NovelloEtAl2000,ObukhovRubilar2002,
SchellstedePerlickLaemmerzahl2016,BandosLechnerSorokinTownsend2020,RussoTownsend2024}.

Previous analyses of Pleba\'nski NLED have shown that gauge-invariant nonlinear
electromagnetic theories may admit nontrivial constant backgrounds compatible with an
effective Hamiltonian bounded from below. The simplest setting in which these magnetic
backgrounds can be displayed explicitly is the single-invariant sector, where the structural
potential depends only on
\[
P=\frac14 P_{\mu\nu}P^{\mu\nu}.
\]
The constrained Hamiltonian analysis of this class was developed in
Ref.~\cite{EscobarPotting2020}, where the effective Hamiltonian governing constant-field
sectors and the corresponding stationarity conditions were obtained. In terms of the
Hamiltonian variables \(\Dvec\) and \(\Hvec\), the invariant is
\begin{equation}
P=\frac12(\Hvec^2-\Dvec^2).
\end{equation}
Writing
\begin{equation}
h=\Hvec^2,
\qquad
d=\Dvec^2,
\end{equation}
the effective Hamiltonian density is
\begin{equation}
\Heff(h,d)
=
-h\Vhat_P(P)+\Vhat(P),
\qquad
P=\frac12(h-d).
\end{equation}
The stationarity conditions must be imposed on the full effective Hamiltonian,
\begin{equation}
\frac{\partial \Heff}{\partial D_i}=0,
\qquad
\frac{\partial \Heff}{\partial H_i}=0.
\end{equation}
The magnetic branch is then obtained by taking
\begin{equation}
\Dvec=0,
\qquad
\Hvec\neq0,
\qquad
P=\frac h2.
\end{equation}
On this branch, the nontrivial stationarity condition reduces to
\begin{equation}
\left(\Vhat_P+h\Vhat_{PP}\right)H_i=0.
\end{equation}
Since \(\Hvec\neq0\), the magnetic stationary background satisfies
\begin{equation}
\Vhat_P(P_0)+h_0\Vhat_{PP}(P_0)=0,
\qquad
P_0=\frac{h_0}{2}.
\end{equation}

On the other hand, previous Hamiltonian analyses already indicated that this magnetic stationary
background is not located at a generic point of the reduced structure~\cite{EscobarPotting2020,EscobarPottingPhysScr2020}. In the Dirac analysis of the
Pleba\'nski first-order theory, the Poisson-bracket matrix among the second-class constraints
has a determinant containing a factor that, in the magnetic branch, is proportional to
\begin{equation}
S_m(h)=\left(\Vhat_P+h\Vhat_{PP}\right)_{P=h/2}.
\end{equation}
The vanishing of this factor makes the constraint matrix degenerate. Remarkably, this is the
same condition that selects the nontrivial magnetic stationary background. Thus, in the
branchwise Hamiltonian formulation, one observes that the magnetic Lorentz-breaking vacuum
lies precisely on a degeneracy surface of the reduced constraint structure.

This coincidence was visible in the reduced Hamiltonian analysis, but its structural origin was
not manifest. In particular, it was not clear whether the degeneracy was a model-dependent
artifact, a peculiarity of the chosen parametrization, or a more fundamental consequence of
the Pleba\'nski constitutive structure. The purpose of the present paper is to identify this
structural mechanism, rather than to construct new symmetry-breaking vacua.

We show that the same Pleba\'nski potential \(V(P,Q)\) that generates the electromagnetic
constitutive relations also determines the magnetic effective Hamiltonian as a complementary
energy at fixed \(\Dvec\). This complementary-energy structure leads to the identity
\begin{equation}
\left.
\frac{\partial \Heff}{\partial H_i}
\right|_{\Dvec}
=
J^{(H)}_{ij}H_j,
\end{equation}
where \(J^{(H)}_{ij}=\partial B_i/\partial H_j|_{\Dvec}\) is the magnetic constitutive
Jacobian. Hence magnetic stationarity with \(\Hvec\neq0\) forces the magnetic order
parameter to become a null direction of \(J^{(H)}\). Since the first-order constitutive
relation enters the Dirac constraint structure, this same Jacobian appears as a local block of the Poisson-bracket matrix among the
second-class constraints. The degeneracy observed in
the Hamiltonian analysis is therefore the constraint-theoretic manifestation of the
constitutive rank loss.

This result motivates the focus on the magnetic branch, but it also raises a natural question:
whether analogous stationary vacua can occur in the other branches of the same
single-invariant theory. We therefore analyze the magnetic, electric, and mixed branches
of the full effective Hamiltonian in the sector \(\Vhat(P)\). In this sector, purely electric
stationary points are obstructed: a configuration that is locally stable within the electric
branch is destabilized once magnetic perturbations of the full effective Hamiltonian are
allowed. Genuine mixed stationary configurations are also nongeneric, since they require
the codimension-two locus
\[
\Vhat_P=\Vhat_{PP}=0,
\]
where the transverse magnetic response vanishes as well. This explains why, within the
single-invariant class considered here, the magnetic branch emerges as the natural candidate
for physically admissible Lorentz-breaking vacua.

The paper is organized as follows. Sections~\ref{sec:constitutive} and \ref{sec:rankloss}
develop the general constitutive mechanism and its rank-loss implication for nontrivial
stationary order parameters. Sections~\ref{sec:plebanski} and \ref{sec:complementary} apply this framework to
Pleba\'nski NLED, showing how the structural potential generates the constitutive
relations and how the effective Hamiltonian becomes the magnetic complementary energy. Section~\ref{sec:constraintlink} explains how the same constitutive rank loss appears in
the Poisson-bracket matrix of second-class constraints, where the degeneracy was originally
detected. Section~\ref{sec:singleinv} specializes the
discussion to single-invariant models and clarifies the status of magnetic, electric, and mixed
branches. Section~\ref{sec:vacuumtheory} discusses the resulting linearized vacuum theory.
Section~\ref{sec:examples} presents explicit two-parameter realizations of the mechanism.
Section~\ref{sec:conclusions} summarizes the main conclusions and outlines physical
implications and future directions. Appendix~\ref{app:Sfactor} derives the corresponding magnetic-response rank-changing factor.

\section{Conservative constitutive systems from physical requirements}
\label{sec:constitutive}

We first isolate the constitutive mechanism that will later be realized in Pleba\'nski
nonlinear electrodynamics. Let \(q^A\), \(A=1,\ldots,N\), denote a set of local constitutive
variables, and let \(p_A(q)\) be the associated response. The reversible work one-form is
\begin{equation}
  \delta W=p_A(q)\,\dd q^A .
\end{equation}
A scalar generating function exists when this one-form is locally exact. This should not be
viewed as a purely formal assumption, but as the local mathematical expression of a
conservative and integrable constitutive response.

The physical requirements are the following. The response is assumed to be local, so that
\(p_A\) depends only on the instantaneous values of the variables \(q^A\). It is also assumed
to be reversible and free of dissipation, so that no net work is produced around an infinitesimal
closed cycle. Finally, we exclude memory and hysteresis, so that the response is a function
of state rather than of the path used to reach that state. In differential form, these
requirements amount to the local integrability condition
\begin{equation}
  \dd(p_A\dd q^A)=0 .
\end{equation}
Equivalently,
\begin{equation}
  \frac{\partial p_A}{\partial q^B}
  =
  \frac{\partial p_B}{\partial q^A}.
\end{equation}
On a simply connected local patch of the constitutive state space, the Poincar\'e lemma
then implies the existence of a scalar function \(F(q)\) such that
\begin{equation}
  \dd F=p_A\dd q^A,
  \qquad
  p_A(q)=\frac{\partial F}{\partial q^A}.
  \label{eq:generatingF}
\end{equation}
Thus \(F\) is the generating potential for the reversible constitutive response.

The corresponding linearized response map is
\begin{equation}
  J_{AB}(q)
  =
  \frac{\partial p_A}{\partial q^B}
  =
  \frac{\partial^2F}{\partial q^A\partial q^B}.
  \label{eq:Jgeneral}
\end{equation}
It is symmetric,
\begin{equation}
  J_{AB}=J_{BA},
\end{equation}
because it is the Hessian of the generating function.

This framework does not apply to dissipative, active, hysteretic, or memory-dependent
systems, for which the response one-form need not be exact. Its relevance for the present
work is that Pleba\'nski nonlinear electrodynamics belongs precisely to this conservative
class: the nonlinear electromagnetic response is generated by the structural potential
\(V(P,Q)\). The rank-loss result derived below is therefore not an additional assumption about Pleba\'nski
theories, but follows from their constitutive integrability together with the nontrivial
stationarity of the corresponding complementary energy.

\section{Nontrivial stationary order parameters and rank loss}
\label{sec:rankloss}

We now show that the existence of a generated constitutive response has a direct consequence
for nontrivial stationary configurations. Given the generating function \(F(q)\), define the
complementary energy
\begin{equation}
  E(q)=q^A p_A(q)-F(q).
  \label{eq:complementary-general}
\end{equation}
Its differential is
\begin{align}
  \dd E
  &= \dd(q^A p_A)-\dd F \nonumber\\
  &= p_A\dd q^A+q^A\dd p_A-p_A\dd q^A \nonumber\\
  &= q^A\dd p_A .
\end{align}
Using
\begin{equation}
  \dd p_A=J_{AB}\dd q^B,
\end{equation}
one obtains
\begin{equation}
  \dd E=q^A J_{AB}\dd q^B ,
\end{equation}
and therefore
\begin{equation}
  \frac{\partial E}{\partial q^B}
  =
  q^A J_{AB}.
  \label{eq:dE-general-index}
\end{equation}
Since the response is conservative, \(J_{AB}=J_{BA}\), this may be written as
\begin{equation}
  \nabla E(q)=J(q)q .
  \label{eq:gradE-Jq}
\end{equation}

Let \(q_0\) be a stationary configuration of the complementary energy,
\begin{equation}
  \nabla E(q_0)=0.
\end{equation}
Equation~\eqref{eq:gradE-Jq} then gives
\begin{equation}
  J(q_0)q_0=0.
  \label{eq:Jqzero}
\end{equation}
If \(q_0=0\), Eq.~\eqref{eq:Jqzero} is an identity and carries no information about the
rank of \(J(q_0)\). If \(q_0\neq0\), however, the stationary configuration itself is a
nonzero null vector of the linearized response map,
\begin{equation}
  q_0\in\kerop J(q_0).
  \label{eq:q0kernel}
\end{equation}
Thus the response map loses rank at such a stationary point. In this sense, for generated
constitutive systems, a nontrivial stationary order parameter is necessarily tied to a
degeneracy of the corresponding linearized response.

This implication should not be read as a universal theorem about spontaneous symmetry
breaking. It applies when the order parameter belongs to a conservative constitutive sector
and when the relevant vacuum condition is the stationarity of the associated complementary
energy. Under these conditions, the order parameter does more than select a direction in field
space: it also selects a null direction of the linearized constitutive response.

A simple radial response illustrates the mechanism. Suppose
\begin{equation}
  p_A(q)=\mu(R)q_A,
  \qquad
  R=q_Aq^A .
\end{equation}
Then
\begin{equation}
  J_{AB}
  =
  \frac{\partial p_A}{\partial q^B}
  =
  \mu(R)\delta_{AB}
  +
  2\mu_R(R)q_Aq_B .
  \label{eq:radial-J}
\end{equation}
For a perturbation \(\xi_\perp^A\) transverse to \(q^A\), namely
\(q_A\xi_\perp^A=0\), the second term in Eq.~\eqref{eq:radial-J} does not contribute.
Hence
\begin{equation}
  J_{AB}\xi_\perp^B=\mu(R)\xi_{\perp A},
\end{equation}
so the transverse eigenvalue is
\begin{equation}
  \lambda_\perp=\mu(R).
\end{equation}
Along the longitudinal direction,
\begin{equation}
  J_{AB}q^B
  =
  \left[\mu(R)+2R\mu_R(R)\right]q_A .
\end{equation}
Thus
\begin{equation}
  \lambda_\parallel
  =
  \mu(R)+2R\mu_R(R).
\end{equation}
A nontrivial stationary point \(q_0\neq0\) therefore satisfies
\begin{equation}
  \lambda_\parallel(R_0)=0.
\end{equation}

The direct response does not become infinite in the order-parameter direction; rather, the
linearized differential response in that direction vanishes. Singular behavior can only arise
if one tries to invert this lower-rank map. This radial example anticipates the
magnetic branch of Pleba\'nski NLED, where the longitudinal magnetic response vanishes
at the Lorentz-breaking vacuum.

\section{Pleba\'nski nonlinear electrodynamics as a generated constitutive theory}
\label{sec:plebanski}

We now apply the preceding constitutive mechanism to nonlinear electrodynamics in the
Pleba\'nski first-order formulation. The independent variables are the gauge potential
\(A_\mu\) and an antisymmetric tensor \(P^{\mu\nu}\). The structural potential depends on
the two Lorentz invariants
\begin{equation}
  P=\frac14 P_{\mu\nu}P^{\mu\nu},
  \qquad
  Q=\frac14 P_{\mu\nu}\widetilde P^{\mu\nu}.
\end{equation}
With the conventions used here, the first-order Lagrangian density may be written as
\begin{equation}
  \widehat{\Lcal}
  =
  -\frac12 P^{\mu\nu}F_{\mu\nu}
  -
  V(P,Q),
  \label{eq:firstorderL}
\end{equation}
where
\begin{equation}
  F_{\mu\nu}=\partial_\mu A_\nu-\partial_\nu A_\mu.
\end{equation}
Overall numerical factors depend on the normalization of \(P^{\mu\nu}\), but they do not
affect the constitutive identities used below.

Varying Eq.~\eqref{eq:firstorderL} with respect to \(P^{\mu\nu}\) gives
\begin{equation}
  F_{\mu\nu}
  =
  -2\frac{\partial V}{\partial P^{\mu\nu}}.
\end{equation}
Indeed,

\begin{equation}
  \delta V
  =
  V_P\delta P+V_Q\delta Q
  =
  \frac12
  \left(
  V_P P_{\mu\nu}
  +
  V_Q\widetilde P_{\mu\nu}
  \right)
  \delta P^{\mu\nu}.
\end{equation}

Hence
\begin{equation}
  \frac{\partial V}{\partial P^{\mu\nu}}
  =
  \frac12
  \left(
  V_P P_{\mu\nu}
  +
  V_Q\widetilde P_{\mu\nu}
  \right),
\end{equation}
and the covariant constitutive relation becomes
\begin{equation}
  F_{\mu\nu}
  =
  -V_P P_{\mu\nu}
  -
  V_Q\widetilde P_{\mu\nu}.
  \label{eq:covconst}
\end{equation}
Thus the structural potential \(V(P,Q)\) is not only part of the first-order dynamics; it is
the covariant generating function of the electromagnetic constitutive relation.

In vector variables we use
\begin{equation}
  P=\frac12(\Hvec^2-\Dvec^2),
  \qquad
  Q=-\Hvec\cdot\Dvec.
  \label{eq:PQvectors}
\end{equation}
Equation~\eqref{eq:covconst} then gives
\begin{equation}
  \begin{pmatrix}
  \Evec\\[2mm]
  \Bvec
  \end{pmatrix}
  =
  \begin{pmatrix}
  -V_P & -V_Q\\[1mm]
  V_Q & -V_P
  \end{pmatrix}
  \begin{pmatrix}
  \Dvec\\[2mm]
  \Hvec
  \end{pmatrix},
  \label{eq:matrixconst}
\end{equation}
or equivalently
\begin{equation}
  \Evec=-V_P\Dvec-V_Q\Hvec,
  \qquad
  \Bvec=V_Q\Dvec-V_P\Hvec.
  \label{eq:EBconst}
\end{equation}

The generating character of \(V(P,Q)\) is also explicit in differential form. From
Eq.~\eqref{eq:PQvectors},
\begin{equation}
  \frac{\partial P}{\partial D_i}=-D_i,
  \qquad
  \frac{\partial Q}{\partial D_i}=-H_i,
  \qquad
  \frac{\partial P}{\partial H_i}=H_i,
  \qquad
  \frac{\partial Q}{\partial H_i}=-D_i.
\end{equation}
Therefore
\begin{equation}
  \frac{\partial V}{\partial D_i}
  =
  -V_PD_i-V_QH_i
  =
  E_i,
  \qquad
  \frac{\partial V}{\partial H_i}
  =
  V_PH_i-V_QD_i
  =
  -B_i.
\end{equation}
Thus
\begin{equation}
  \dd V
  =
  \Evec\cdot\dd\Dvec
  -
  \Bvec\cdot\dd\Hvec.
  \label{eq:dV}
\end{equation}

In the notation of Sec.~\ref{sec:constitutive}, the complete electromagnetic constitutive
system is therefore identified as
\begin{equation}
  q^A=(D_i,H_i),
  \qquad
  p_A=(E_i,-B_i),
  \qquad
  F(q)=V(P,Q).
\end{equation}
This is the precise sense in which the Pleba\'nski formulation realizes a conservative
generated constitutive structure. The structural potential \(V(P,Q)\) generates the response
map, while the effective Hamiltonian that selects the vacua will be identified as the
corresponding complementary energy in the magnetic sector.

\section{The effective Hamiltonian as magnetic complementary energy}
\label{sec:complementary}

The structural potential \(V(P,Q)\) generates the constitutive relation, but the physical
vacua are not obtained by minimizing \(V\) itself. In the constrained Hamiltonian analysis
of Pleba\'nski nonlinear electrodynamics, after implementing the second-class constraints,
the effective Hamiltonian density governing the constant-field sectors is~\cite{EscobarPotting2020}
\begin{equation}
  \Heff
  =
  -\Hvec^2V_P-QV_Q+V(P,Q).
  \label{eq:Heffgeneral}
\end{equation}
This distinction is essential: \(V(P,Q)\) controls the constitutive response, whereas
\(\Heff\) controls the vacuum structure.

The bridge between these two objects is that \(\Heff\) can be interpreted as a complementary
energy for the magnetic constitutive sector. Holding \(\Dvec\) fixed, define the partial
magnetic generating function
\begin{equation}
  F_{\Dvec}(\Hvec)
  =
  -V(P,Q).
\end{equation}
Using Eq.~\eqref{eq:dV} at fixed \(\Dvec\), one obtains
\begin{equation}
  \left.\frac{\partial F_{\Dvec}}{\partial H_i}\right|_{\Dvec}
  =
  -\left.\frac{\partial V}{\partial H_i}\right|_{\Dvec}
  =
  B_i .
\end{equation}
Therefore \(F_{\Dvec}\) generates the magnetic response at fixed \(\Dvec\),
\begin{equation}
  \Bvec
  =
  \frac{\partial F_{\Dvec}}{\partial\Hvec}.
\end{equation}
The corresponding complementary energy is
\begin{equation}
  E_{\Dvec}(\Hvec)
  =
  H_iB_i-F_{\Dvec}(\Hvec)
  =
  \Hvec\cdot\Bvec+V(P,Q).
\end{equation}
Since
\begin{equation}
  \Hvec\cdot\Bvec
  =
  \Hvec\cdot(V_Q\Dvec-V_P\Hvec)
  =
  -QV_Q-\Hvec^2V_P,
\end{equation}
we find
\begin{equation}
  E_{\Dvec}(\Hvec)
  =
  -\Hvec^2V_P-QV_Q+V(P,Q)
  =
  \Heff .
  \label{eq:HeffComplementary}
\end{equation}

Although the complementary-energy construction is performed at fixed \(\Dvec\), the
identity \(\Heff(\Dvec,\Hvec)=E_{\Dvec}(\Hvec)\) is pointwise on the full constitutive
state space. The phrase ``at fixed \(\Dvec\)'' only specifies that the complementary
transformation is taken with respect to the magnetic pair \((\Hvec,\Bvec)\), while
\(\Dvec\) is treated as a parameter. Thus \(\Heff\) remains a function of both
\(\Dvec\) and \(\Hvec\), and the full vacuum conditions still require stationarity with
respect to both variables.

This observation is the central conceptual point. It explains why the stationarity of
\(\Heff\) is tied to the degeneracy of the magnetic constitutive map: both the magnetic
response and the effective Hamiltonian derive from the same structural potential \(V(P,Q)\).

Since \(F_{\Dvec}\) generates the magnetic response, its Hessian is the magnetic
constitutive Jacobian
\begin{equation}
  J^{(H)}_{ij}
  =
  \left.\frac{\partial B_i}{\partial H_j}\right|_{\Dvec}
  =
  \frac{\partial^2 F_{\Dvec}}{\partial H_i\partial H_j}.
\end{equation}

Because \(\Heff=E_{\Dvec}=H_iB_i-F_{\Dvec}\), its derivative at fixed \(\Dvec\) is

\begin{align}
  \frac{\partial \Heff}{\partial H_i}
  &=
  \frac{\partial}{\partial H_i}
  \left(H_jB_j-F_{\Dvec}\right)  \nonumber\\
  &=
  B_i
  +
  H_j
  \frac{\partial B_j}{\partial H_i}
  -
  \frac{\partial F_{\Dvec}}{\partial H_i}  \nonumber\\
  &=
  H_jJ^{(H)}_{ji}.
\end{align}

Since \(J^{(H)}_{ij}=J^{(H)}_{ji}\), this gives
\begin{equation}
  \left.
  \frac{\partial \Heff}{\partial H_i}
  \right|_{\Dvec}
  =
  J^{(H)}_{ij}H_j.
  \label{eq:Heff-JH}
\end{equation}

Therefore, any configuration satisfying the magnetic stationarity condition with a nonzero
magnetic component, \(\Hvec_0\neq0\), satisfies
\begin{equation}
  J^{(H)}_{ij}(\Dvec_0,\Hvec_0)H_{0j}=0.
\end{equation}

Thus such a configuration lies on a surface where the magnetic linearized map
\(\delta\Hvec\mapsto\delta\Bvec\), at fixed \(\Dvec\), loses rank.

\section{Constitutive rank loss and the second-class constraint matrix}
\label{sec:constraintlink}

The previous section identified the constitutive origin of the lower-rank magnetic vacuum:
by the complementary-energy identity, a nontrivial magnetic stationary configuration makes
the magnetic order parameter a null direction of the magnetic constitutive Jacobian
\(J^{(H)}\). We now explain why this same loss of rank is detected in the Hamiltonian
constraint analysis as a degeneracy of the Poisson-bracket matrix among the second-class
constraints.

The key point is that, in the first-order Pleba\'nski formulation, the constitutive relations
are not external relations imposed after the Hamiltonian analysis. They enter the constraint
structure itself. This can be seen explicitly in the Dirac analysis of
Ref.~\cite{EscobarPotting2020}. Since \(P^{\mu\nu}\) is treated as an independent
first-order field, the spatial components \(P^{ij}\), which encode the magnetic auxiliary
variables, have canonical momenta \(\pi_{ij}\) that vanish as primary constraints. In the
notation of Ref.~\cite{EscobarPotting2020}, this primary constraint is
\begin{equation}
  \Theta^5_{ij}
  =
  \Delta^4_{ij}
  =
  \pi_{ij}
  \approx0 .
  \label{eq:Theta5}
\end{equation}
Preserving this constraint in time produces the spatial part of the covariant constitutive
relation, namely
\begin{equation}
  \Theta^6_{ij}
  =
  (\partial_iA_j-\partial_jA_i)
  +
  V_P P_{ij}
  +
  V_Q\widetilde P_{ij}
  \approx0 .
  \label{eq:Theta6}
\end{equation}
This is precisely the \(ij\)-component of
\begin{equation}
  F_{\mu\nu}
  +
  V_P P_{\mu\nu}
  +
  V_Q\widetilde P_{\mu\nu}
  =
  0,
  \label{eq:cov-constitutive-constraint}
\end{equation}
which is the Pleba\'nski constitutive relation.

In vector notation, Eq.~\eqref{eq:Theta6} is equivalent, up to conventional signs
associated with dualizing spatial antisymmetric tensors, to the magnetic constitutive
condition
\begin{equation}
  \Bvec
  =
  \boldsymbol{\mathcal B}(\Dvec,\Hvec),
  \qquad
  \mathcal B_i(\Dvec,\Hvec)
  =
  V_QD_i-V_PH_i .
  \label{eq:Bmap-constraintlink}
\end{equation}
Thus, locally and up to nonsingular redefinitions of the constraint basis, the magnetic
part of the second-class constraint set contains the algebraic information
\begin{equation}
  B_i[A]-\mathcal B_i(\Dvec,\Hvec)
  \approx0 .
  \label{eq:Psi-vector-constraint}
\end{equation}
This is not an additional assumption; it is the vector form of the spatial constitutive
constraint \(\Theta^6_{ij}\).

The primary constraint paired with this constitutive condition is the momentum conjugate
to the magnetic auxiliary variable. In tensor notation this is
\(\Theta^5_{ij}=\pi_{ij}\approx0\). Since \(\Hvec\) is obtained from the spatial
components \(P^{ij}\) by dualization, the vector form of the momentum conjugate to
\(H_i\) is correspondingly the dual of \(\pi_{ij}\). We denote it schematically by
\begin{equation}
  \Pi^{(H)}_i
  \sim
  \frac12\epsilon_{ijk}\pi_{jk}.
\end{equation}
The precise sign and normalization depend on the convention used to identify \(P^{ij}\)
with \(\Hvec\), but this does not affect the rank argument.

The local Poisson-bracket block between \(\Theta^5_{ij}\) and \(\Theta^6_{kl}\), or
equivalently between \(\Pi^{(H)}_i\) and the vector form of the magnetic constitutive
constraint, then contains the derivative of the constitutive map:
\begin{equation}
  \left\{
  \Pi^{(H)}_i(x),
  B_j[A](y)-\mathcal B_j(\Dvec,\Hvec)(y)
  \right\}
  =
  -
  \frac{\partial\mathcal B_j}{\partial H_i}
  \delta^{(3)}(x-y),
  \label{eq:PB-JH-block}
\end{equation}
up to conventional signs and normalization factors. Since
\begin{equation}
  \frac{\partial\mathcal B_j}{\partial H_i}
  =
  \left.
  \frac{\partial B_j}{\partial H_i}
  \right|_{\Dvec}
  =
  J^{(H)}_{ji},
  \label{eq:JH-block-identification}
\end{equation}
the magnetic constitutive Jacobian appears, up to conventional signs and nonsingular
changes of constraint basis, as a local algebraic block of the Poisson-bracket matrix among
the second-class constraints. This is the precise point at which the constitutive loss of rank
is inherited by the Dirac analysis.

The same structure is visible in the determinant of the constraint matrix. In
Ref.~\cite{EscobarPotting2020}, the determinant of the Poisson-bracket matrix of the
second-class constraints \(\Theta^a\), \(a=3,4,5,6\), was found to be proportional to
\begin{equation}
  V_P^2S^2 .
  \label{eq:Dirac-det-S}
\end{equation}
The factor \(S\) is therefore the nontrivial rank-changing factor of the reduced constraint
structure. Its appearance is not independent of the constitutive response: it reflects the fact
that the constitutive Jacobian enters the local Poisson-bracket matrix through the block
associated with \(\Theta^5_{ij}\) and \(\Theta^6_{ij}\).

Thus the surface that was originally detected through the degeneracy of the Poisson-bracket
matrix among the second-class constraints has a constitutive origin. The Pleba\'nski first-order formalism forces the constitutive
relation to enter the constraint structure, and the magnetic Jacobian \(J^{(H)}\), which
controls the linearized map \(\delta\Hvec\mapsto\delta\Bvec\), appears inside the
corresponding local block of the Poisson-bracket matrix. Therefore, when the magnetic
stationarity condition makes this Jacobian lose rank, the Poisson-bracket matrix of
second-class constraints becomes degenerate on the same surface.

In the next section we specialize this mechanism to the single-invariant sector. The purpose
is no longer to identify where the degeneracy appears in the Dirac matrix, but to analyze
which stationary branches of the full effective Hamiltonian can realize Lorentz-breaking
backgrounds.

\section{Vacuum branches in the single-invariant sector}
\label{sec:singleinv}

In this section we
specialize to the single-invariant sector and analyze the stationary branches of the full
effective Hamiltonian. This is important because the magnetic branch is not imposed from
the outset: the stationarity conditions must first be written for the full Hamiltonian and
only then restricted to magnetic, electric, or mixed configurations.

We consider
\begin{equation}
  V(P,Q)=\Vhat(P).
\end{equation}
In this case, Eq.~\eqref{eq:Heffgeneral} reduces to
\begin{equation}
  \Heff(h,d)
  =
  -h\Vhat_P(P)
  +
  \Vhat(P),
  \qquad
  P=\frac12(h-d),
  \label{eq:Heffhat}
\end{equation}
where
\begin{equation}
  h=\Hvec^2,
  \qquad
  d=\Dvec^2 .
\end{equation}
Since
\begin{equation}
  \frac{\partial P}{\partial H_i}=H_i,
  \qquad
  \frac{\partial P}{\partial D_i}=-D_i,
\end{equation}
the full stationarity equations are
\begin{equation}
  \left.
  \frac{\partial\Heff}{\partial H_i}
  \right|_{\Dvec}
  =
  -
  \left(
  \Vhat_P+h\Vhat_{PP}
  \right)H_i,
  \label{eq:dHeffdH-single}
\end{equation}
and
\begin{equation}
  \left.
  \frac{\partial\Heff}{\partial D_i}
  \right|_{\Hvec}
  =
  \left(
  h\Vhat_{PP}
  -
  \Vhat_P
  \right)D_i .
  \label{eq:dHeffdD-single}
\end{equation}
We now analyze the magnetic, electric, and mixed branches of these same equations.
\subsection{Magnetic branch}
\label{subsec:magnetic-branch}

The purely magnetic branch is defined by
\begin{equation}
  \Dvec=0,
  \qquad
  \Hvec\neq0,
  \qquad
  P=\frac h2 .
\end{equation}
On this branch, Eq.~\eqref{eq:dHeffdD-single} is automatically satisfied, while
Eq.~\eqref{eq:dHeffdH-single} gives
\begin{equation}
  \Vhat_P(P_0)+h_0\Vhat_{PP}(P_0)=0,
  \qquad
  P_0=\frac{h_0}{2},
  \qquad
  h_0>0 .
  \label{eq:singlecondition}
\end{equation}
It is convenient to introduce the magnetic branch factor
\begin{equation}
  S_m(h):=
  \left(\Vhat_P+h\Vhat_{PP}\right)_{P=h/2}.
  \label{eq:Smdef}
\end{equation}
Thus the nontrivial magnetic stationarity condition is simply
\begin{equation}
  S_m(h_0)=0.
\end{equation}

The magnetic constitutive relation in the single-invariant sector is
\begin{equation}
  \Bvec=-\Vhat_P(P)\Hvec .
  \label{eq:Bhatmag}
\end{equation}
At fixed \(\Dvec\), the corresponding magnetic constitutive Jacobian is
\begin{equation}
  J^{(H)}_{ij}
  =
  \left.
  \frac{\partial B_i}{\partial H_j}
  \right|_{\Dvec}
  =
  -\left(\Vhat_P\delta_{ij}+\Vhat_{PP}H_iH_j\right).
  \label{eq:JHhat}
\end{equation}
For perturbations transverse to \(\Hvec\), the eigenvalue is
\begin{equation}
  \lambda_\perp=-\Vhat_P,
  \label{eq:lamperp}
\end{equation}
whereas along the longitudinal direction \(\Hvec\) one obtains
\begin{equation}
  \lambda_\parallel
  =
  -\left(\Vhat_P+h\Vhat_{PP}\right).
  \label{eq:lamparallel}
\end{equation}
After restricting to the magnetic branch,
\begin{equation}
  \lambda_\parallel=-S_m(h).
\end{equation}
Therefore
\begin{equation}
  S_m(h_0)=0
  \quad\Longleftrightarrow\quad
  \lambda_\parallel(h_0)=0 .
  \label{eq:mainequivalence}
\end{equation}
Thus the magnetic Lorentz-breaking vacuum is a lower-rank constitutive vacuum. The
rank loss is longitudinal: the response along the order-parameter direction vanishes at
first order, while the transverse response remains regular provided
\begin{equation}
  \Vhat_P(P_0)\neq0 .
\end{equation}

A constant magnetic background \(\Hvec_0\neq0\) selects a preferred spatial direction and
therefore breaks the Lorentz group to the subgroup that preserves the corresponding
magnetic two-form. For a purely magnetic background chosen along a fixed spatial axis, the
residual symmetry is \(SO(1,1)\times SO(2)\), corresponding to boosts along the magnetic
direction and rotations around it. In the Pleba\'nski Hamiltonian realization of such a vacuum,
vacuum selection and longitudinal rank loss are controlled by the same condition.

\subsection{Electric branch}
\label{subsec:electric-branch}

We next consider the purely electric branch,
\begin{equation}
  \Hvec=0,
  \qquad
  \Dvec\neq0,
  \qquad
  P_0=-\frac{d_0}{2}<0,
  \qquad
  d_0=\Dvec^2 .
\end{equation}
On this branch, Eq.~\eqref{eq:dHeffdH-single} is automatically satisfied, while
Eq.~\eqref{eq:dHeffdD-single} gives
\begin{equation}
  \Vhat_P(P_0)=0 .
  \label{eq:electric_stationarity}
\end{equation}
If one restricts attention to the electric branch alone, local radial stability would require
\begin{equation}
  \Vhat_{PP}(P_0)>0 .
\end{equation}
However, this does not define a minimum of the full effective Hamiltonian.

To see this, keep \(d=d_0\) fixed and turn on a small magnetic perturbation
\begin{equation}
  h=\Hvec^2>0.
\end{equation}
Then
\begin{equation}
  P=P_0+\frac h2 .
\end{equation}

Using \(\Vhat_P(P_0)=0\), the effective Hamiltonian expands as
\begin{equation}
  \Heff(h,d_0)
  =
  \Heff(0,d_0)
  -
  \frac38\,\Vhat_{PP}(P_0)h^2
  +
  O(h^3).
  \label{eq:electric_instability}
\end{equation}
This instability is invisible at quadratic order in the magnetic variables. Indeed,
differentiating Eq.~\eqref{eq:dHeffdH-single} once more with respect to \(H_j\), one finds
\begin{equation}
  \left.
  \frac{\partial^2\Heff}{\partial H_i\partial H_j}
  \right|_{\Hvec=0}
  =
  -\Vhat_P(P_0)\delta_{ij}
  =
  0,
\end{equation}
where the last equality follows from the electric stationarity condition
\(\Vhat_P(P_0)=0\). The first nonvanishing magnetic variation therefore appears at order
\(h^2=(\Hvec^2)^2\), namely at quartic order in \(\Hvec\).

Therefore, if \(\Vhat_{PP}(P_0)>0\), the energy decreases under arbitrarily small magnetic
perturbations. The electric stationary point is then unstable in the full theory. Thus, in the
single-invariant sector, a purely electric stationary point that is locally stable within the
electric branch is destabilized by magnetic perturbations of the full effective Hamiltonian.

\subsection{Mixed branch}
\label{subsec:mixed-branch}

Finally, consider a stationary configuration with both electric and magnetic components,
\begin{equation}
  \Hvec\neq0,
  \qquad
  \Dvec\neq0 .
\end{equation}
Equations~\eqref{eq:dHeffdH-single} and \eqref{eq:dHeffdD-single} then imply
\begin{equation}
  \Vhat_P+h\Vhat_{PP}=0,
  \qquad
  h\Vhat_{PP}-\Vhat_P=0 .
\end{equation}
Since \(h\neq0\), these two equations are equivalent to
\begin{equation}
  \Vhat_P(P_0)=0,
  \qquad
  \Vhat_{PP}(P_0)=0 .
  \label{eq:mixed_codim2}
\end{equation}
Thus, in a single-invariant Pleba\'nski theory, a stationary configuration with both electric
and magnetic components is not generic. It can occur only at a codimension-two locus of the
structural potential.

At such a point the transverse magnetic eigenvalue also vanishes,
\begin{equation}
  \lambda_\perp=-\Vhat_P=0.
\end{equation}
The mixed stationary configuration is therefore more degenerate than the magnetic lower-rank
vacua discussed above, where only the longitudinal magnetic eigenvalue vanishes while
\begin{equation}
  \Vhat_P(P_0)\neq0 .
\end{equation}
Consequently, a mixed branch in the single-invariant sector is not a robust analogue of the
magnetic Lorentz-breaking vacuum unless a separate higher-order stability analysis is
performed.

Combining the previous results, the magnetic branch emerges as the natural candidate for
physically admissible Lorentz-breaking vacua in the single-invariant sector: purely electric
stationary points are destabilized by magnetic perturbations, while genuine mixed stationary
points require a stronger codimension-two degeneracy.

\section{Linearized vacuum theory and vacuum restriction}
\label{sec:vacuumtheory}

The lower-rank character of the magnetic vacuum should not be interpreted as an infinite
physical response. It means that the direct linearized constitutive map has a nontrivial
kernel, and therefore its inverse is not defined on the full perturbation space.

Let the magnetic vacuum be

\begin{equation}
  \Dvec_0=0,
  \qquad
  \Hvec_0=H_0\nvec,
  \qquad
  H_0=\sqrt{h_0},
  \qquad
  \nvec^2=1 .
\end{equation}

A magnetic fluctuation can be decomposed as
\begin{equation}
  \delta\Hvec
  =
  \delta H_\parallel \nvec
  +
  \delta\Hvec_\perp,
  \qquad
  \nvec\cdot\delta\Hvec_\perp=0 .
\end{equation}
Equivalently, introducing the longitudinal and transverse projectors
\begin{equation}
  (P_\parallel)_{ij}=n_i n_j,
  \qquad
  (P_\perp)_{ij}=\delta_{ij}-n_i n_j ,
\end{equation}
one has
\begin{equation}
  \delta\Hvec_\parallel
  =
  \delta H_\parallel\nvec
  =
  P_\parallel\delta\Hvec,
  \qquad
  \delta\Hvec_\perp=P_\perp\delta\Hvec .
\end{equation}

At the magnetic vacuum, the magnetic constitutive Jacobian takes the spectral form
\begin{equation}
  J^{(H)}_0
  =
  \lambda_\perp(P_0)P_\perp
  +
  \lambda_\parallel(P_0)P_\parallel .
\end{equation}
Using Eqs.~\eqref{eq:lamperp} and \eqref{eq:lamparallel}, together with the vacuum condition
\(S_m(h_0)=0\), this becomes
\begin{equation}
  J^{(H)}_0
  =
  -\Vhat_P(P_0)P_\perp .
  \label{eq:JH-vacuum-projector}
\end{equation}
Therefore
\begin{equation}
  \delta\Bvec
  =
  J^{(H)}_0\delta\Hvec
  =
  -\Vhat_P(P_0)\delta\Hvec_\perp .
  \label{eq:deltaB-vacuum}
\end{equation}
The longitudinal perturbation does not produce a first-order magnetic response,
\begin{equation}
  \delta\Bvec_\parallel=0,
\end{equation}
whereas the transverse response remains regular provided
\begin{equation}
  \Vhat_P(P_0)\neq0 .
\end{equation}
Equivalently,
\begin{equation}
  \ker J^{(H)}_0=\mathrm{span}\{\nvec\},
  \qquad
  \mathrm{Im}\,J^{(H)}_0=\{\delta\Bvec\,:\,\nvec\cdot\delta\Bvec=0\}.
\end{equation}

Thus a prescribed linearized magnetic response \(\delta\Bvec\) is compatible with the vacuum
linearization only if
\begin{equation}
  \nvec\cdot\delta\Bvec=0.
  \label{eq:magnetic-compatibility}
\end{equation}
This is the concrete compatibility condition associated with the lower-rank magnetic response.
It replaces the longitudinal inverse-response equation that would exist away from the vacuum.

Indeed, along the regular magnetic branch, where \(S_m(h)\neq0\), one may formally invert
the magnetic constitutive map as

\begin{equation}
  \delta\Hvec
  =
  -\frac{1}{\Vhat_P}\delta\Bvec_\perp
  -
  \frac{1}{S_m(h)}\delta\Bvec_\parallel .
  \label{eq:formal-inverse}
\end{equation}
This formula is not valid at the vacuum. Taking \(S_m(h)\to0\) after solving the regular
inverse problem produces an apparent divergence if \(\delta\Bvec_\parallel\neq0\). The
vacuum theory is instead obtained by imposing the vacuum condition first, in which case
\(\delta\Bvec_\parallel\) is not an allowed response and Eq.~\eqref{eq:magnetic-compatibility}
must be imposed.

This also clarifies the dynamical interpretation of the rank-changing surface. Away from the
vacuum, the regular branch may be evolved using the inverse response map. If a regular
trajectory approaches \(S_m=0\), however, the inverse description ceases to be valid. A smooth approach to the lower-rank surface is possible only when the limiting linearized
data satisfy the compatibility conditions of the degenerate map. Otherwise the regular reduced evolution
becomes singular before it can be continued through the surface.

Thus \(S_m=0\) should not be interpreted, by itself, as an inconsistency of the underlying
first-order theory. Rather, it is a rank-changing stratum where the regular reduced
Hamiltonian description changes character. Evolution on that stratum must be formulated
with the lower-rank constraints imposed from the start, or with a separate Dirac analysis
adapted to the degenerate surface. Whether a given solution reaches this surface in finite time,
approaches it asymptotically, or is dynamically excluded is a question about the full initial-value
problem and not about the local constitutive identity alone.

This lower-rank behavior is confined to the magnetic longitudinal sector. In the full
single-invariant constitutive response, the electric displacement block remains regular at the
magnetic vacuum:
\begin{equation}
  \delta\Evec
  =
  -\Vhat_P(P_0)\delta\Dvec ,
\end{equation}
again assuming \(\Vhat_P(P_0)\neq0\). The rank loss discussed here is therefore the
longitudinal magnetic rank loss identified in Sec.~\ref{subsec:magnetic-branch}.

There is a further, stronger construction: one may restrict the theory to remain on the
vacuum manifold itself. In the magnetic branch this local vacuum manifold is
\begin{equation}
  \mathcal M_{\rm vac}
  =
  \left\{
  (\Dvec,\Hvec):\,
  \Dvec=0,\quad \Hvec^2=h_0
  \right\}.
  \label{eq:vacuum-manifold}
\end{equation}
Equivalently, one may parametrize the magnetic variable as
\begin{equation}
  \Hvec(x)=H_0\nvec(x),
  \qquad
  H_0=\sqrt{h_0},
  \qquad
  \nvec^2(x)=1.
  \label{eq:H-vacuum-param}
\end{equation}
The local constitutive variable on the restricted vacuum manifold is therefore not the magnitude
of \(\Hvec\), but only its orientation.

Linearizing the constraints that define \(\mathcal M_{\rm vac}\) gives
\begin{equation}
  \delta\Dvec=0,
  \qquad
  \delta(\Hvec^2)=2\Hvec_0\cdot\delta\Hvec=0.
\end{equation}
Thus
\begin{equation}
  \Hvec_0\cdot\delta\Hvec=0,
  \qquad
  \delta H_\parallel=0.
  \label{eq:radial-removed}
\end{equation}
The radial fluctuation is precisely the longitudinal direction of the ambient magnetic
constitutive map. Hence the direction in which \(J^{(H)}_0\) loses rank is not an
independent fluctuation once the vacuum constraint is imposed from the start.

Using the parametrization \(\Hvec=H_0\nvec\), an allowed vacuum-restricted fluctuation has
the form
\begin{equation}
  \delta\Hvec=H_0\,\delta\nvec,
  \qquad
  \nvec\cdot\delta\nvec=0.
\end{equation}
Since, at the magnetic vacuum,
\begin{equation}
  \Bvec=-\Vhat_P(P_0)\Hvec,
\end{equation}
one obtains
\begin{equation}
  \delta\Bvec
  =
  -\Vhat_P(P_0)\delta\Hvec
  =
  -\Vhat_P(P_0)H_0\,\delta\nvec .
  \label{eq:vacuum-restricted-deltaB}
\end{equation}
Thus the vacuum-restricted constitutive response is regular on the tangent space of the
vacuum manifold, provided
\begin{equation}
  \Vhat_P(P_0)\neq0.
\end{equation}
The singular longitudinal inverse never appears, because the longitudinal fluctuation has
already been removed by the constraint \(\Hvec^2=h_0\).

It is useful to distinguish these two linearizations. The ambient linearized theory around the
vacuum is a lower-rank constitutive system: arbitrary \(\delta\Hvec\) may be considered, but
the allowed responses must satisfy the compatibility condition
\(\nvec\cdot\delta\Bvec=0\). The strictly vacuum-restricted theory is instead the projected
theory on \(T_{\Hvec_0}\mathcal M_{\rm vac}\), where the radial degenerate direction has
already been removed. Thus imposing the vacuum from the start is not equivalent to first
solving the regular inverse-response equations and then taking \(S_m\to0\); the vacuum
constraint removes the problematic longitudinal direction before any inversion is attempted.

\section{Concrete two-parameter models}
\label{sec:examples}

The constitutive mechanism derived above is independent of the detailed functional form of
the single-invariant potential. In particular, the lower-rank condition is not a consequence of
global boundedness of the effective Hamiltonian. It follows locally from the magnetic
stationarity condition and from the fact that the same structural potential generates both the
magnetic constitutive response and the magnetic complementary energy.

As shown above, any nontrivial magnetic stationary point on the magnetic branch satisfies
\(S_m(h_0)=0\), and hence \(\lambda_\parallel(h_0)=0\). This statement is local and constitutive: no assumption about the global boundedness of \(\Heff\) is needed for this
algebraic conclusion.

The role of the explicit models summarized in Table~\ref{tab:models1} is therefore more
specific. The potentials, admissible parameter regions, and vacuum magnitudes listed in the
table are taken from the two-parameter constructions analyzed in
Ref.~\cite{PlacidoFloresEtAl2026}. They show that the mechanism is realized in concrete
single-invariant Pleba\'nski families for which the additional physical requirements of
boundedness, local stability, and magnetic spontaneous Lorentz symmetry breaking can be
satisfied simultaneously. These requirements are model dependent, whereas the lower-rank
character of the magnetic stationary point is not.

\begin{table}[htbp]
\centering
\small
\setlength{\tabcolsep}{5pt}
\renewcommand{\arraystretch}{1.22}
\begin{tabular}{
|>{\centering\arraybackslash}p{0.28\textwidth}
|>{\centering\arraybackslash}p{0.28\textwidth}
|>{\centering\arraybackslash}p{0.36\textwidth}|
}
\hline
Potential \(\Vhat(P)\) &
Admissible magnetic region &
Vacuum magnitude \(h_0\) \\
\hline

\(\displaystyle
-P+\frac{\lambda P^3}{1+\eta P^2}
\)
&
\(\displaystyle
\frac98\lambda\leq \eta<
\left(
\frac{25}{16}+\frac{5\sqrt5}{16}
\right)\lambda
\)
&
\(\displaystyle
\begin{gathered}
h_0\ \text{is the positive real solution of}\\
(\eta^3-\eta^2\lambda)h_0^6
+12\eta^2h_0^4\\
+(48\eta-240\lambda)h_0^2+64=0
\end{gathered}
\)
\\
\hline

\(\displaystyle
-P-\frac{\lambda}{\eta}\ln(1+\eta P)
\)
&
\(\displaystyle
\lambda>8,\qquad \eta>0
\)
&
\(\displaystyle
h_0=
\frac{\lambda-2+\sqrt{\lambda(\lambda-8)}}{\eta}
\)
\\
\hline

\(\displaystyle
-P-\frac{\lambda}{\eta}\left(1-e^{-\eta P}\right)
\)
&
\(\displaystyle
\lambda>\frac{e^{3/2}}{2},
\qquad
\eta>0
\)
&
\(\displaystyle
h_0=
\frac{
1-2W_{-1}\!\left(-\frac{e^{1/2}}{2\lambda}\right)
}{\eta}
\)
\\
\hline

\end{tabular}
\caption{
Representative single-invariant Pleba\'nski potentials admitting stable magnetic
Lorentz-breaking vacua in appropriate parameter regions, following the two-parameter
models analyzed in Ref.~\cite{PlacidoFloresEtAl2026}. The admissible regions are model
dependent and encode the coexistence of boundedness, local stability, and magnetic SSB.
The lower-rank condition is model independent: whenever the magnetic stationary point
exists, \(S_m(h_0)=0\) implies \(\lambda_\parallel(h_0)=0\). Here \(W_{-1}\) denotes the
\(-1\) branch of the Lambert \(W\) function.
}
\label{tab:models1}
\end{table}

Thus the examples separate two logically distinct issues. The existence of a physically
admissible magnetic vacuum depends on global and local properties of the chosen potential,
such as boundedness of \(\Heff\) and positivity of the relevant Hessian directions. By contrast,
the lower-rank nature of that magnetic stationary point follows from the constitutive identity
alone. The table therefore illustrates that stable magnetic Lorentz-breaking vacua can occur,
while the lower-rank nature of such vacua is a model-independent consequence of the Pleba\'nski
constitutive structure.

\section{Conclusions and outlook}
\label{sec:conclusions}

Previous Hamiltonian analyses had already shown that the magnetic Lorentz-breaking vacua of
these Pleba\'nski theories lie on surfaces where the determinant of the Poisson-bracket matrix
among the second-class constraints vanishes. The main result of this work is to identify the
structural origin of that coincidence. It is not an accidental feature of the explicit models, nor
merely a peculiarity of the reduced Hamiltonian parametrization. It follows from the fact that
the same structural potential that generates the magnetic constitutive response also determines
the effective Hamiltonian through the corresponding magnetic complementary energy. Moreover,
because the spatial constitutive relation enters the second-class constraint set, the magnetic
constitutive Jacobian appears as a local block of the Dirac matrix. Consequently,
\begin{equation}
  \left.
  \frac{\partial \Heff}{\partial H_i}
  \right|_{\Dvec}
  =
  J^{(H)}_{ij}H_j .
\end{equation}
Thus a nontrivial magnetic stationary point necessarily lies on a surface where the linearized
magnetic response loses rank; through the corresponding block in the Dirac matrix, the
Poisson-bracket matrix of second-class constraints becomes degenerate on the same surface.

In the single-invariant sector \(V(P,Q)=\Vhat(P)\), this result reduces on the magnetic branch to
\begin{equation}
  S_m(h_0)=0
  \quad\Longleftrightarrow\quad
  \Vhat_P(P_0)+h_0\Vhat_{PP}(P_0)=0
  \quad\Longleftrightarrow\quad
  \lambda_\parallel(h_0)=0,
  \qquad
  P_0=\frac{h_0}{2}.
\end{equation}
The same condition that selects the nontrivial magnetic background therefore makes the
longitudinal magnetic response vanish. This explains why magnetic Lorentz-breaking vacua
found in the branchwise Hamiltonian analysis coincide with degeneracy surfaces of the
reduced constraint structure.

This conclusion is local and constitutive. It does not rely on the effective Hamiltonian being
globally bounded from below. Boundedness and local stability are additional, model-dependent
requirements needed to interpret a stationary point as a physically admissible vacuum. The
two-parameter examples summarized in Table~\ref{tab:models1} show that these additional
requirements can be satisfied, while the lower-rank nature of the magnetic stationary point
follows from the constitutive identity itself.

This places Pleba\'nski NLED alongside bumblebee-type vector models in the broader
landscape of constrained nonlinear field theories with spontaneous Lorentz breaking, where
the Hamiltonian and constraint structure can determine the physically admissible vacuum
beyond what is implied by the Lagrangian potential alone.

The lower-rank interpretation also fixes how perturbations should be treated. Apparent
divergences arise only if one tries to invert the regular response map and then take
\(S_m\to0\). At the vacuum, the correct linearized theory is instead a lower-rank system with
compatibility conditions on the allowed responses. If the theory is restricted to the vacuum
manifold, the radial degenerate fluctuation is removed from the start and only tangent
orientation modes remain.

Several directions remain open. The analysis should be extended to general two-invariant
potentials \(V(P,Q)\), where mixed or dyonic vacua require a coupled stationarity and stability
analysis. It would also be useful to perform the Dirac constraint analysis directly on the
degenerate surface, rather than as a limit of the regular branch. Finally, applications to
Casimir-type geometries, magnetic black-hole backgrounds, and causality/characteristic
propagation should be revisited using the lower-rank vacuum theory.

\appendix

\section{Degeneracy factor for the magnetic response}
\label{app:Sfactor}

In this appendix we derive the rank-changing factor associated with the magnetic constitutive
Jacobian for a general two-invariant structural potential \(V(P,Q)\). This calculation is
independent of the full Dirac matrix determinant, but it shows explicitly that, on the magnetic
branch, the constitutive rank-changing factor reduces to the same combination that appears in
the Hamiltonian degeneracy condition. We use the conventions
of the main text,
\begin{equation}
  P=\frac12(\Hvec^2-\Dvec^2),
  \qquad
  Q=-\Hvec\cdot\Dvec,
\end{equation}
and the constitutive relation
\begin{equation}
  \Bvec=V_Q\Dvec-V_P\Hvec .
\end{equation}
At fixed \(\Dvec\), the magnetic constitutive Jacobian is
\begin{equation}
  J^{(H)}_{ij}
  =
  \left.
  \frac{\partial B_i}{\partial H_j}
  \right|_{\Dvec}.
\end{equation}
Introducing the abbreviations
\begin{equation}
  a=V_P,
  \qquad
  b=V_{PP},
  \qquad
  c=V_{PQ},
  \qquad
  e=V_{QQ},
\end{equation}
one obtains
\begin{equation}
  J^{(H)}
  =
  -a\,I
  -
  b\,\Hvec\Hvec^T
  +
  c\left(\Hvec\Dvec^T+\Dvec\Hvec^T\right)
  -
  e\,\Dvec\Dvec^T .
  \label{eq:JH-general-appendix}
\end{equation}
This follows directly from
\begin{equation}
  \frac{\partial P}{\partial H_j}=H_j,
  \qquad
  \frac{\partial Q}{\partial H_j}=-D_j .
\end{equation}

The matrix \(J^{(H)}\) acts trivially on the direction orthogonal to the plane spanned by
\(\Hvec\) and \(\Dvec\). The corresponding eigenvalue is
\begin{equation}
  \lambda_\perp=-a=-V_P .
\end{equation}
The nontrivial rank-changing information is therefore contained in the two-dimensional block
on the plane generated by \(\Hvec\) and \(\Dvec\).

Using the matrix determinant lemma, Eq.~\eqref{eq:JH-general-appendix} gives
\begin{equation}
  \det J^{(H)}
  =
  -V_P\,S_H,
  \label{eq:detJH-S}
\end{equation}
where
\begin{equation}
\begin{aligned}
  S_H
  &=
  \left(V_{PP}V_{QQ}-V_{PQ}^{\,2}\right)
  \left(\Hvec^2\Dvec^2-Q^2\right)
  \\
  &\quad
  +
  V_P
  \left(
  \Dvec^2 V_{QQ}
  +
  \Hvec^2 V_{PP}
  +
  2Q V_{PQ}
  \right)
  +
  V_P^2 .
\end{aligned}
\label{eq:Sfactor-general}
\end{equation}
Thus \(S_H=0\) is the nontrivial rank-changing surface of the magnetic response whenever
\(V_P\neq0\).

In the magnetic branch,
\begin{equation}
  \Dvec=0,
  \qquad
  Q=0,
  \qquad
  h=\Hvec^2,
\end{equation}
Eq.~\eqref{eq:Sfactor-general} reduces to
\begin{equation}
  S_H
  =
  V_P\left(V_P+hV_{PP}\right).
\end{equation}
If \(V_P\neq0\), the nontrivial degeneracy condition is therefore
\begin{equation}
  V_P+hV_{PP}=0.
\end{equation}
For a single-invariant theory \(V(P,Q)=\Vhat(P)\), evaluated on the magnetic branch
\(P=h/2\), this becomes
\begin{equation}
  \left(\Vhat_P+h\Vhat_{PP}\right)_{P=h/2}=0.
\end{equation}
This is precisely the magnetic stationarity condition and, equivalently, the vanishing of the
longitudinal magnetic response eigenvalue,
\begin{equation}
  \lambda_\parallel
  =
  -\left(\Vhat_P+h\Vhat_{PP}\right).
\end{equation}

\end{document}